\documentclass[sigconf, screen]{acmart}

\usepackage{xcolor} 
\usepackage{array} 

\newcommand{\litbadge}[1]{%
  \fcolorbox{purple}{purple!10}{\textsf{\textbf{\footnotesize#1}}}%
}
\newcommand{\onlinebadge}[1]{%
  \fcolorbox{blue}{blue!10}{\textsf{\textbf{\footnotesize#1}}}%
}

\newcolumntype{L}[1]{>{\raggedright\arraybackslash}p{#1}}

\AtBeginDocument{%
  }

\copyrightyear{2025}
\acmYear{2025}
\setcopyright{cc}
\setcctype{by}
\acmConference[AutomotiveUI '25]{17th International Conference on Automotive User Interfaces and Interactive Vehicular Applications}{September 21--25, 2025}{Brisbane, QLD, Australia}
\acmBooktitle{17th International Conference on Automotive User Interfaces and Interactive Vehicular Applications (AutomotiveUI '25), September 21--25, 2025, Brisbane, QLD, Australia}
\acmDOI{10.1145/3744333.3747834}
\acmISBN{979-8-4007-2013-0/2025/09}
\begin{document}

\title{Animal Interaction with Autonomous Mobility Systems: Designing for Multi-Species Coexistence}

\author{Tram Thi Minh Tran}
\email{tram.tran@sydney.edu.au}
\orcid{0000-0002-4958-2465}
\affiliation{Design Lab, Sydney School of Architecture, Design and Planning,
  \institution{The University of Sydney}
  \city{Sydney}
  \state{NSW}
  \country{Australia}
}

\author{Xinyan Yu}
\email{xinyan.yu@sydney.edu.au}
\orcid{0000-0001-8299-3381}
\affiliation{Design Lab, Sydney School of Architecture, Design and Planning,
  \institution{The University of Sydney}
  \city{Sydney}
  \state{NSW}
  \country{Australia}
}

\author{Marius Hoggenmueller}
\email{marius.hoggenmueller@sydney.edu.au}
\orcid{0000-0002-8893-5729}
\affiliation{Design Lab, Sydney School of Architecture, Design and Planning
  \institution{The University of Sydney} 
  \city{Sydney}
  \state{NSW}
  \country{Australia}
}

\author{Callum Parker}
\email{callum.parker@sydney.edu.au}
\orcid{0000-0002-2173-9213}
\affiliation{Design Lab, Sydney School of Architecture, Design and Planning,
  \institution{The University of Sydney}
  \city{Sydney}
  \state{NSW}
  \country{Australia}
}

\author{Paul Schmitt}
\email{Pauls@massrobotics.org}
\orcid{0000-0001-8237-648X}
\affiliation{
  \institution{MassRobotics}
  \city{Boston}
  \state{Massachusetts}
  \country{United States}
}

\author{Julie Stephany Berrio Perez}
\email{stephany.berrioperez@sydney.edu.au}
\orcid{0000-0003-3126-7042}
\affiliation{The Australian Centre for Robotics,
  \institution{The University of Sydney}
  \city{Sydney}
  \state{NSW}
  \country{Australia}
}

\author{Stewart Worrall}
\email{stewart.worrall@sydney.edu.au}
\orcid{0000-0001-7940-4742}
\affiliation{The Australian Centre for Robotics,
  \institution{The University of Sydney}
  \city{Sydney}
  \state{NSW}
  \country{Australia}
}

\author{Martin Tomitsch}
\email{Martin.Tomitsch@uts.edu.au}
\orcid{0000-0003-1998-2975}
\affiliation{Transdisciplinary School,
  \institution{University of Technology Sydney}
  \city{Sydney}
  \state{NSW}
  \country{Australia}
}

\renewcommand{\shortauthors}{Tran et al.}

\begin{abstract}

Autonomous mobility systems increasingly operate in environments shared with animals, from urban pets to wildlife. However, their design has largely focused on human interaction, with limited understanding of how non-human species perceive, respond to, or are affected by these systems. Motivated by research in Animal-Computer Interaction (ACI) and more-than-human design, this study investigates animal interactions with autonomous mobility through a multi-method approach combining a scoping review (45 articles), online ethnography (39 YouTube videos and 11 Reddit discussions), and expert interviews (8 participants). Our analysis surfaces five key areas of concern: Physical Impact (e.g., collisions, failures to detect), Behavioural Effects (e.g., avoidance, stress), Accessibility Concerns (particularly for service animals), Ethics and Regulations, and Urban Disturbance. We conclude with design and policy directions aimed at supporting multispecies coexistence in the age of autonomous systems. This work underscores the importance of incorporating non-human perspectives to ensure safer, more inclusive futures for all species.

\end{abstract}

\ccsdesc[500]{Human-centered computing~Interaction design}

\keywords{animals, wildlife, automated vehicles, autonomous vehicles, interaction}

\maketitle

\section{Introduction}

Autonomous mobility systems, including autonomous vehicles (AVs)~\cite{brown2023halting, manger2024reality}, drones~\cite{lingam2024challenges}, and mobile robots~\cite{yu2024understanding, pelikan2024encountering, shin2024delivering}, are becoming increasingly present in both urban and rural environments. A growing body of work in Human–Computer Interaction (HCI) has explored how people interact with these technologies, and how these technologies operate within and reshape infrastructure. However, this work has largely prioritised human needs and perspectives, leaving a limited understanding of how non-human animals experience these systems. For example, a keyword search for `animals' in the AutomotiveUI proceedings returns only papers where animals are mentioned briefly or in the context of anthropomorphism\footnote{Search conducted on 31 March 2025 via the ACM Digital Library: \url{https://dl.acm.org/action/doSearch?AllField=animals&expand=all&ConceptID=119411}. The query returned 764 results.}. 

Emerging strands of research have begun to address this gap. Studies have explored wildlife detection~\cite{li2024endangered}, documented animal stress responses~\cite{rebolo2019drones}, and raised ethical questions about how autonomous systems should treat animals~\cite{mancini2023responsible, hagendorff2023speciesist}. However, these efforts remain fragmented, and much remains unknown about how animals perceive, respond to, and are affected by different autonomous systems in shared environments. These systems differ not only in form and function but also in how they move through space: drones navigate aerial environments, mobile robots operate at ground level in pedestrian areas, and AVs travel on roads at higher speeds; each presenting distinct challenges and sensory signals for animals. Responding to recent calls in the Animal-Computer Interaction (ACI) research agenda~\cite{mancini2023responsible} \textit{`to focus on the risks that technology poses for animals and on the impacts of technology that does not necessarily target them,'} and building on more-than-human design perspectives that challenge anthropocentric assumptions and recognise non-human agency~\cite{giaccardi2020technology, heidegger1977question, frauenberger2019entanglement, hauser2018design, wakkary2021things, tomitsch2023designing}, this study seeks to reframe animals not merely as edge cases or obstacles, but as relational actors within shared environments, whose presence and behaviours influence how autonomous systems operate and should be designed.

We adopt a multi-method approach to develop a holistic understanding of animal interaction with autonomous mobility systems. A scoping review examines how animals appear in the academic literature. An online ethnography of user-generated YouTube videos and Reddit discussions offers insights into real-world encounters. Finally, expert interviews with ecologists, animal behaviourists, and mobility and ACI researchers respond to both the literature and real-world findings, contributing integrated insights to inform more animal-aware design and policymaking. Our research is guided by the following research questions (RQs):

\begin{itemize} 
\item \textbf{RQ1}: How do animals perceive, respond to, and become affected by autonomous mobility systems across different species and contexts? 
\item \textbf{RQ2}: What design and policy opportunities exist to better support multispecies coexistence with autonomous mobility systems? 
\end{itemize}

The paper makes two contributions: (1) A thematic synthesis of literature and real-world evidence on how autonomous mobility systems impact animals across species and contexts. (2) Design and policy directions for supporting multispecies co-existence with autonomous systems, based on this synthesis and enriched by expert insights. In doing so, we draw attention to the need for greater inclusion of non-human perspectives in autonomous mobility research and encourage the HCI and mobility research communities to engage more deeply with multispecies coexistence.

\enlargethispage{\baselineskip}

\section{Related Work}

\subsection{More Than Human Perspective and Animal-Computer Interaction}

Human-centred design has long shaped HCI by prioritising usability and placing the human at the centre of interaction~\cite{cooper2004inmates}. While influential, this anthropocentric paradigm has faced growing critique for marginalising other beings and ecologies. \citet{giaccardi2020technology} argue that it treats technology as a \textit{`standing reserve'}~\cite{heidegger1977question}—a resource optimised for one-to-one, human-serving interactions. \citet{frauenberger2019entanglement} similarly calls for a shift towards relational ontologies and ethical accountability, as emerging technologies challenge the field’s ontological foundations. \citet{hauser2018design}, drawing on post-phenomenology, advocates for \textit{`decentring the human,'} building on Verbeek’s proposition to move beyond asking what technology can do for users toward asking \textit{`how human beings can be present in the world and how the world can be present for human beings'}~\cite{verbeek2015beyond}. \citet{roudavski2025dingo} advance this agenda by integrating ecocentric and technocentric perspectives into a more-than-human framework for decision-making, recognising the creative agency of both human and nonhuman actors in shaping environmental governance and design.

Within this space, ACI has emerged as a sister field that focuses on the design and use of technology with, for, and by animals. ACI explores both the risks that technologies pose to animals and the opportunities to design technologies that benefit them, advocating for approaches that go beyond human-centred paradigms~\cite{mancini2013animal, mancini2023responsible}. Recent work in ACI spans topics such as animal communication, behaviour recognition, enrichment, and support for working animals; for example, auditory enrichment systems to support elephant welfare in zoos~\cite{mastali2024play}, or haptic interfaces that facilitate communication with detection dogs in distracting environments~\cite{hopper2024towards}. While many of these systems are tailored to specific species and contexts, emerging AI-enabled technologies, such as autonomous mobility systems in urban and rural environments, raise broader concerns about how animals are represented, sensed, and responded to. \citet{hagendorff2023speciesist} note that many AI models are trained on datasets in which speciesist patterns prevail, risking the misrepresentation or exclusion of non-human agency. Scholars have called for expanding AI ethics to include animals, ensuring that automated decision-making supports rather than marginalises non-human species~\cite{singer2023ai}.

\subsection{Animal Interactions with Autonomous Mobility}  

Autonomous mobility systems are technologies capable of navigating and acting within physical environments with varying degrees of independence, often without direct human control. These systems operate across diverse domains (e.g., roads, shared spaces, the air, or natural landscapes) and differ in how they move, sense, and signal their presence. Their growing deployment raises important questions about how animals perceive and respond to these technologies.

Insights from road ecology offer a useful starting point: roads and vehicles have long been known to pose risks to animals, including habitat fragmentation, barrier effects, and vehicle-related mortality~\cite{forman2003road}. Animal crossing behaviour is influenced by traffic speed, noise, and lighting, with species-specific variations in avoidance and adaptation~\cite{nag2020darwin}. Compared to traditional vehicles, AVs introduce new variables that may alter animal responses. For example, electric AVs pose challenges for noise-reliant species that depend on auditory cues to navigate~\cite{hwang2024towards}. More generally, \citet{bendel2018towards} highlights that semi-autonomous and autonomous machines operate across diverse environments, frequently encountering pets, urban-dwelling animals, and wildlife, and notes that such encounters often leave  \textit{`[these] sensitive creatures and capable of suffering, [...] disturbed, irritated and scared, or even injured and killed.'}

Recent progress in animal–autonomous mobility research has spanned technical, ethical, and empirical domains. Technically, advances in perception systems and simulators have improved detection and avoidance capabilities~\cite{rosique2019systematic, li2024endangered}. Ethically, debates have emerged around how systems should respond to animals in unavoidable collisions, how animal life is valued in automated decision-making, and how speciesist bias in AI models may reinforce anthropocentric assumptions~\cite{mancini2023responsible, hagendorff2023speciesist}. Empirical studies have also begun to examine the impact of specific technologies, such as drones, on animal behaviour~\cite{rebolo2019drones}. However, despite these developments, current efforts remain scattered across disciplines and autonomous system types, with limited synthesis. Much less is known about how animals perceive, respond to, or are affected by autonomous systems from a more holistic and relational perspective. Therefore, our study adopts a broader view by examining how various systems interact with animals across shared environments. It brings a more-than-human perspective into autonomous mobility research, foregrounding animals as actors within these multispecies contexts.

\section{Methods}

\begin{figure*} [ht]
    \centering
\includegraphics[width=0.86\linewidth]{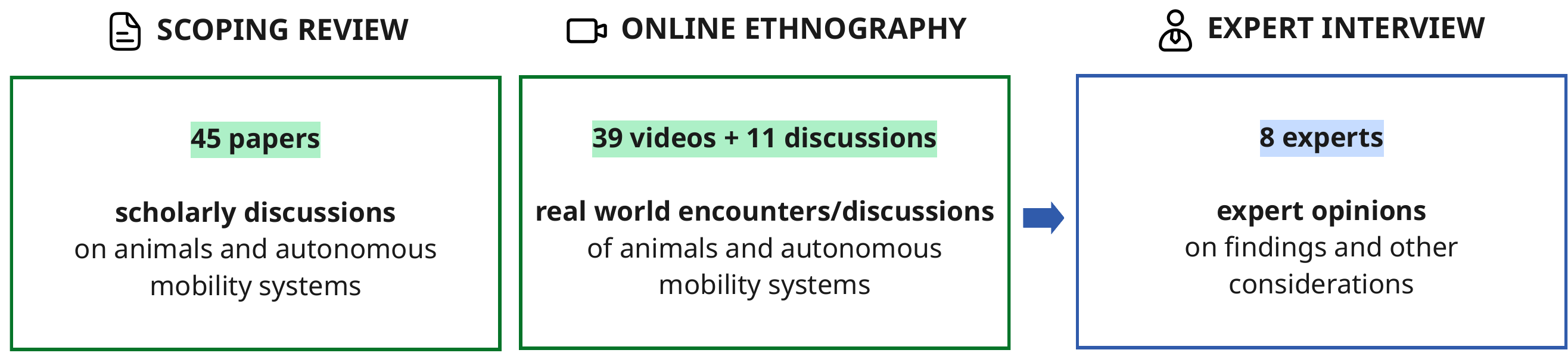}
    \caption{Overview of the multi-method approach used in this study.}
    \Description{This figure presents a three-part diagram summarising the study’s multi-method approach. From left to right, it includes: (1) a scoping review with “45 papers” highlighting scholarly discussions on animals and autonomous mobility systems, (2) an online ethnography featuring “39 videos + 11 discussions” capturing real-world encounters and public discourse, and (3) expert interviews with “8 experts” providing opinions on findings and additional considerations. Arrows connect the components, illustrating how the study builds from literature to real-world observations to expert interpretation.}
    \label{fig:methods}
\end{figure*}

This study employs a multi-method approach to ensure a comprehensive understanding of animal interaction with autonomous mobility systems (see \autoref{fig:methods}). The \textit{scoping review} maps existing research and identifies key themes, while the \textit{online ethnography} captures real-world encounters and discussion topics not well documented in the academic literature. Finally, \textit{expert interviews} help interpret the findings of both methods and uncover additional considerations related to technology, ecology, and policy.

\subsection{Scoping Review}

\paragraph{Databases} Our search used the ACM Digital Library, which includes proceedings from the AutomotiveUI, CHI, ACI, and HRI conferences, all of which are highly relevant due to their focus on autonomous mobility systems, interactive technologies, animal-computer interaction and human-robot interaction. Additionally, we used Google Scholar to capture papers not indexed in this database.

\paragraph{Search Queries} To systematically identify relevant studies, we structured the search terms into two categories: animal-related terms (e.g., animal, wildlife) and autonomous mobility terms (e.g., autonomous vehicles, drones, delivery robots). These keywords were combined in different Boolean search queries (see \autoref{appendix:scoping}). The search returned 1,231 results (ACM: 805; Google Scholar: 681), with \textbf{933} unique papers after removing duplicates.

\paragraph{Inclusion Criteria}

\begin{itemize}
    \item Discussed animal interactions with autonomous mobility and factors influencing animal behaviour.
    \item Mentioned design considerations for integrating these systems into environments shared with animals.
    \item Contributed theoretical, speculative, or ethical discussions relevant to animal interactions. 
\end{itemize}

\paragraph{Exclusion Criteria}

\begin{itemize}
    \item Focused on animal-inspired designs for human use or made incidental references to animals.
    \item Focused solely on technical aspects (e.g., detection algorithms).
    \item Examined technology for wildlife conservation, including drones designed primarily for animal monitoring, conservation, or smart farming. 
    \item Not a primary research contribution, including proceedings collections, demonstrations, courses, tutorials, interviews, and editorials.
\end{itemize}

\paragraph{Screening Process:} The first author screened the identified papers based on the title, abstract, and a keyword search (Ctrl+F) for animal-related terms within the full text (resulting in \textbf{40 papers}). An additional \textbf{7 papers} were identified through backward snowballing, examining the reference lists of the selected papers. In the next phase, the first and second authors reviewed the full text to determine exclusions. In parallel, both researchers began noting topical keywords for each paper, an early form of descriptive coding that informed the later thematic analysis. During this stage, several decisions were made regarding the inclusion of specific robot types. Lawn mower robots were included as they operate in semi-public green spaces, encountering wildlife and domestic animals. Quadruped robots (robotic dogs), envisioned as service dogs or delivery assistants, were likewise considered relevant. This resulted in a total of \textbf{45 papers} for synthesis.

\paragraph{Data Analysis}

Relevant excerpts from each paper were extracted into a spreadsheet and initially charted using descriptive codes: species type (e.g., kangaroo, deer, dog), autonomous system type (e.g., AVs, drones, delivery robots), and study type (empirical or non-empirical). The first author then conducted an inductive thematic analysis to identify recurring narratives across the literature, which were subsequently grouped under five thematic categories: Physical Impact, Behavioural Effects, Ethics and Regulations, Accessibility Concerns, and Urban Disturbance.

\subsection{Online Ethnography}

\paragraph{Platforms} We combined visual evidence and discussion-based insights. Specifically, YouTube provided video footage of encounters, while Reddit hosted discussions in communities such as r/SelfDrivingCars and r/drones, where users shared first-hand experiences and concerns. TikTok was excluded because its content largely overlapped with YouTube and was dominated by AI-generated videos (e.g., cats `driving' cars).

\paragraph{Search Queries} Our search strategy followed a similar keyword structure to the scoping review but incorporated more specific terms (e.g., deer, cat, dog) and an additional brand-specific category (e.g., Waymo, Tesla) (see \autoref{appendix:ethnography}). All searches were conducted without logging into accounts to minimise algorithmic bias stemming from personalised viewing history. This process returned a total of \textbf{60} YouTube videos and \textbf{29} Reddit discussions.

\paragraph{Screening Process} The screening process was slightly adapted to align with the characteristics of each platform. \textbf{YouTube}: The first author conducted the initial search and identified videos featuring animal-autonomous system encounters. The first and second authors then reviewed the full videos to determine inclusion based on observed patterns in the footage (using the criteria below). A total of \textbf{39} videos were retained. \textbf{Reddit}: We focused on original posts\footnote{In the context of Reddit, the `original post' refers to the initial submission made by a user to start a new discussion. It typically includes a title and a main body of text where the user introduces a topic, asks a question, shares an experience, or presents information.} discussing topics related to animals and autonomous systems (e.g., incidents, safety concerns, ethical debates, and unexpected interactions). This approach ensured that the study captured the core discussion topics without being skewed by individual debates or unrelated conversations. 
The final dataset included \textbf{11} discussions.

\paragraph{Inclusion Criteria}
\begin{itemize}
    \item Instances of autonomous systems encountering animals, including cases where the system successfully detects and avoids animals or results in a collision.
    \item Animals inside moving vehicles without an accompanying human.
    \item News reports that provide direct footage of the interaction.
\end{itemize}

\paragraph{Exclusion Criteria}
\begin{itemize}
    \item Company-produced content (e.g., promotional material from Waymo, Tesla).
    \item Staged testing scenarios that do not reflect spontaneous real-world encounters.
    \item Repeated content (e.g., multiple posts covering the same incident). \end{itemize}


\paragraph{Data Analysis}


The first author analysed selected YouTube videos and Reddit posts involving animals and autonomous systems. Similar to the literature data, content was categorised by species type and autonomous system type. While initial themes were developed inductively, we then applied the same five thematic categories from the scoping review to support consistency across both datasets and enable comparative analysis.

\subsection{Expert Interviews}

\paragraph{Expert Profiles}
Experts were identified through the scoping review and the authors' professional networks. 
They included ecologists, animal behaviourists, and researchers in autonomous mobility and ACI. A total of \textbf{eight experts} were interviewed (see \autoref{tab:expert_profiles} in Appendix for detailed profiles).

Expert interviews were conducted via Zoom between 24 March and 9 April 2025 by either the first or second author (except for the first interview, which was jointly conducted to ensure a consistent protocol). Each session lasted between 30–45 minutes. The interviews followed a three-part structure: (1) experts introduced their background and any prior work related to animals; (2) a general discussion explored animal interactions with autonomous systems; and (3) experts responded to selected findings from the scoping review and online ethnography. 

\paragraph{Data Analysis}
The audio recordings were transcribed using Otter.ai and validated by the interviewers. Inductive thematic analysis~\cite{braun2006thematic} was performed by the first author, with themes discussed and refined collaboratively with other authors.

\section{Results}


\subsection{Scoping Review and Online Ethnography}
\label{sec:lit_ethno}

This subsection presents the findings from both the scoping review \litbadge{Literature} and online ethnography \onlinebadge{Ethnography} using a shared thematic structure developed during the analysis: Physical Impact, Behavioural Effects, Ethics and Regulations, Accessibility Concerns, and Urban Disturbance. Before presenting the thematic findings in detail, we first provide an overview of the datasets by reporting the distribution of autonomous system types and species types across both sources, followed by a breakdown of study types within the scoping review.

\textbf{Autonomous mobility systems}: were grouped based on form, function, and operating context: road-based vehicles, mobile robots, aerial systems, maintenance robots, and other/generic technologies (\autoref{tab:autonomous_system_types}).

\textbf{Species types}: were based on species referenced where identifiable (\autoref{tab:animal_categories}). In the reviewed papers, animals were sometimes discussed generically, using broad terms such as animals or wildlife, without reference to specific species or groups.



\textbf{Study types} (scoping review only): Of the 45 papers, 9 (20\%) were empirical studies that directly investigated animal interactions with autonomous mobility systems. These studies spanned a variety of technologies and contexts, including drone-wildlife interactions, dog responses to delivery robots and drones, accessibility challenges involving guide dogs, and hedgehog safety in relation to lawn mower robots. The remaining 36 (80\%) were discussion-based or mention-only papers, where animals featured in broader debates on ethics, policy, or ACI directions.

\begin{table*}[ht]
\centering
\small
\caption{Autonomous system types and their characteristics in relation to animal interaction.}
\begin{tabular}{p{2.5cm} L{2.8cm} p{1.4cm} p{2.1cm} L{4.6cm}}
\toprule
\textbf{System Type} & \textbf{Examples} & \textbf{Literature (n=45)} & \textbf{YouTube+Reddit (n=50)} & \textbf{Key Characteristics\newline (drawn from literature)} \\
\midrule
Road-based vehicles & AVs, buses, rideshare & 15 (33\%) & 26 (53\%) & Large, often operate at higher speeds. \\
Mobile robots & Delivery robots, service robots, robotic dogs & 9 (20\%) & 14 (29\%) & Operate in shared pedestrian spaces, close proximity. \\
Aerial systems & Drones, robotic bats & 15 (33\%) & 10 (20\%) & Airborne, generate noise and vibration. \\
Maintenance robots & Lawn mower robots & 2 (5\%) & - & Operate in semi-public green spaces, low to ground. \\
Other/Generic & Mobility-assisted tech, generic robots & 4 (9\%) & - & Generic or conceptual systems, offer broad insights on animal interactions. \\
\bottomrule
\end{tabular}
\label{tab:autonomous_system_types}
\end{table*}

\begin{table*}[ht]
\centering
\small
\caption{Species represented in literature and online data, with counts and percentages.}
\begin{tabular}{L{3.6cm} L{5.8cm} p{2cm} p{2.2cm}}
\toprule
\textbf{Species Type} & \textbf{Examples} & \textbf{Literature (n=33)*} & \textbf{YouTube+Reddit (n=76)*} \\
\midrule
Domestic animals & Dogs, Guide/Service dogs, Police dogs, Cats, Pets & 19 (58\%) & 31 (41\%) \\
Wild ungulates & Deer, Moose, Cows & 5 (15\%) & 16 (21\%) \\
Large birds \& birds of prey & Birds, Eagles, Geese, Ducks, Hawks, Magpies, Pigeons, Ravens, Waterfowl, Osprey & 3 (9\%) & 12 (16\%) \\
Large predators & Bears, Lions, Tigers, Cheetah & 1 (3\%) & 5 (7\%) \\
Aquatic predators & Alligators, Crocodiles, Sharks & - & 6 (8\%) \\
Primates & Chimpanzees & - & 1 (1\%) \\
Large non-predator mammals & Horses, Giraffes, Elephants, Rams, Kangaroos & 2 (6\%) & 5 (7\%) \\
Small or unusual species & Hedgehogs & 3 (9\%) & - \\
\bottomrule
\addlinespace
  \multicolumn{4}{p{14cm}}{\textsuperscript{*}Numbers reflect instances of each species type. Some papers, videos, and discussions referenced multiple species, therefore totals may exceed the number of sources.}
\end{tabular}
\label{tab:animal_categories}
\end{table*}


\begin{figure*} [ht]
    \centering
\includegraphics[width=1\linewidth]{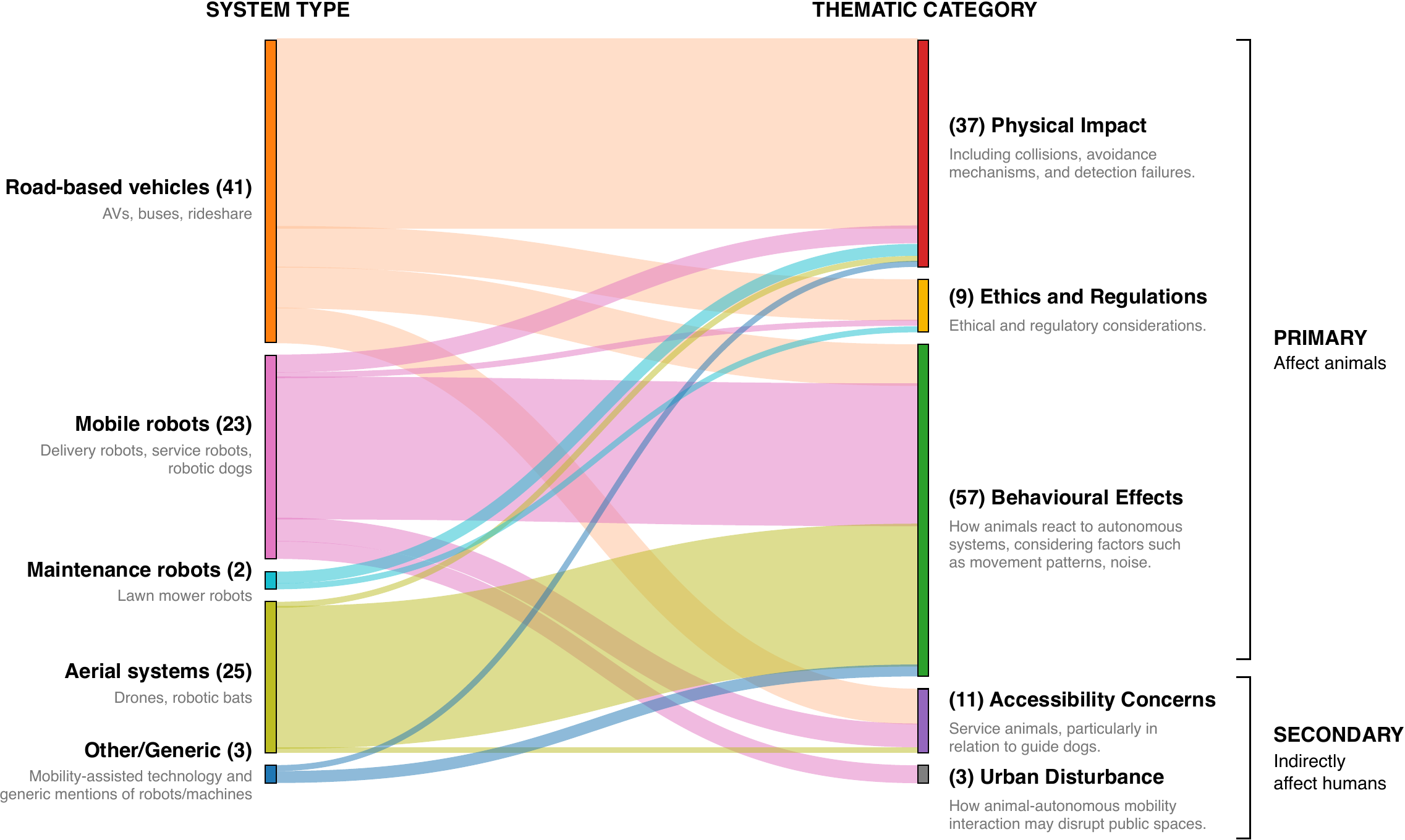}
    \caption{Sankey diagram mapping the relationships between autonomous system types and the five thematic categories. Primary impacts directly affect animals, while secondary impacts indirectly affect humans through their relationship with animals. The aim of the figure is to visualise the distribution and intensity of concerns across system types, helping to identify which technologies are associated with which types of impacts.}
    \Description{This Sankey diagram illustrates the connections between five types of autonomous systems (left) and five thematic impact categories (right). System types include road-based vehicles (41 instances), mobile robots (23), maintenance robots (2), aerial systems (25), and other or generic mentions (3). These are linked to thematic categories: Physical Impact (37), Ethics and Regulations (9), Behavioural Effects (57), Accessibility Concerns (11), and Urban Disturbance (3). The diagram distinguishes between primary impacts (those that directly affect animals) and secondary impacts (those that indirectly affect humans via animal interactions). The width of each flow indicates the frequency of concern across system-impact pairings.}
    \label{fig:sankey-review}
\end{figure*}


\subsubsection{Physical Impact} \hfill\\
\indent \litbadge{Literature} \textit{Risk of injury or death caused by autonomous systems and responses aimed at preventing such incidents.}

One of the most frequently addressed topics in the literature is the direct physical impact of autonomous systems on animals. Various studies emphasise the risk of injury or death when animals enter environments shared with AVs or autonomous machines. In particular, smart cities, despite their promise of efficiency and safety, continue to pose threats to animals, especially vehicles that cannot detect small or fast-moving species~\cite{thrift2021killer, mancini2023responsible, kalik2007automotive}. In road settings, wildlife such as deer, bears, and ducks are commonly cited as sources of unpredictable encounters, with human drivers and AV operators alike must react quickly to avoid harm~\cite{akridge2024bus}. 

Beyond conceptual discussions, empirical studies offer concrete evidence of harm caused by autonomous systems. For instance, \citet{rasmussen2021wildlife} experimentally demonstrate that many robotic lawn mowers fail to detect hedgehogs, resulting in significant physical damage, especially to small or juvenile individuals. Moreover, the threat extends beyond ground-based systems. Research on drones highlights similar risks, including injuries to birds and the disruption of sensitive wildlife habitats~\cite{chang2017spiders}.

In response to these risks, a range of interventions and mitigation strategies have been proposed or implemented. For example, \citet{parks2019mediating} noted that Volvo’s Large Animal Detection System uses radar and machine learning to detect and respond to animals based on their size, proximity, and movement, while Mercedes vehicles have incorporated motion-detecting headlights and infrared cameras to improve night-time visibility of large animals. More forward-looking proposals include a wildlife priority mode for AVs~\cite{li2024creating}, and drone-specific visual or audio repellents~\cite{chang2017spiders}. 

\onlinebadge{Ethnography} \textit{Successful/unsuccessful detections and avoidance; inconsistent visual representations of animals.}

Among the 39 analysed videos, 15 involved road-based AVs, primarily Tesla and Waymo. Successful detections were more common. Several Tesla FSD\footnote{Tesla's Full Self-Driving (FSD) is a suite of advanced driver-assistance features that includes Autopilot capabilities.} videos showed the system slowing, rerouting, or braking in response to deer, bears, or smaller animals. Waymo also appeared to reliably detect dogs, whether accompanied by owners or roaming alone. However, ambiguous or failed detections were also observed. Collisions were documented in two Tesla FSD videos in which stationary deer were not detected at night. In some cases, animals were visualised inconsistently or incorrectly; for instance, a deer was shown as a small bird, a large cat, or as a pedestrian symbol on the Tesla in-vehicle display. Waymo visualised dogs as white moving dots. One Tesla driver reported that the system failed to detect two dogs on the roadside, prompting manual disengagement. In another case, a Tesla navigated around a dead animal that did not appear on the display, raising questions about how detection logic operates in the absence of visual confirmation.

Reddit posts echoed similar cases. One post recounted a Tesla striking a dog while in FSD mode, with the driver claiming that the system showed no reaction. In another widely discussed case, a Waymo vehicle fatally collided with a small dog despite the system having detected the animal. The spokesperson noted that the dog's trajectory made avoidance difficult, and neither the AV nor the safety driver braked. Others revisited long-standing issues, like the `kangaroo problem,'~\cite{Deahl2017} where AV systems struggle to predict the animal movement. A particularly detailed post argued that avoiding deer collisions may be virtually impossible due to their erratic behaviour, highlighting deeper design challenges for AVs in dynamic wildlife environments. In contrast, some posts highlighted successful detections, such as a Waymo stopping for a cat. 

\subsubsection{Behavioural Effects} \hfill\\
\indent \litbadge{Literature} \textit{Animal stress and behavioural disruption from AVs, robots, and drones; emphasis on species-aware design and habituation.}

In the context of road-based vehicles, semi-autonomous and autonomous systems operating in various environments can disturb or displace domestic and wild animals~\cite{bendel2018towards}. Transport networks, including those supporting AVs, contribute to habitat degradation by impairing animals’ ability to communicate, attract mates, or avoid predators~\cite{thrift2021killer}. Noise, in particular, is a key environmental stressor and can be harmful if levels are $\geq$ 85 dB~\cite{jin2020acoussist}. Some species are also sensitive to high-frequency sounds. For example, dogs and bats respond to ultrasound frequencies~\cite{yaghoubisharif2022mobile}. The shift to quieter electric and hybrid vehicles introduces new challenges, especially for animals that rely on auditory cues to navigate traffic environments. Guide dogs, for instance, have been reported to struggle in judging the speed and proximity of silent vehicles~\cite{hwang2024towards}. Finally, design considerations for AVs often overlook how animal behaviour shapes system interaction. As noted by \cite{joshi2023autonomous}, animals are among the `external road users' whose reactions influence how an AV should behave, particularly in proximity-based encounters. However, most AV communication systems are currently geared toward human interpretation. 

Unlike AVs, mobile robots operate in closer proximity to animals, particularly pets and urban wildlife, making their behavioural presence more immediate and disruptive. Studies consistently report uncomfortable or adverse reactions from dogs, who often interpret these robots as novel and unpredictable stimuli. In an empirical study with over 200 dog owners, \citet{vaataja2023exploring} found that while 61\% of the dogs exhibited neutral behaviour on first encounter, 8\% avoided and 9\% reacted aggressively to delivery robots. Notably, many of the dogs that responded negatively did not habituate to repeated encounters. Dogs with pre-existing anxiety or phobic conditions, such as deprivation syndrome, were more likely to exhibit avoidance, freezing, or fight-or-flight responses when confronted with robots~\cite{zamansky2018effects}. Looking ahead, \citet{nijholt2020virtual} imagines scenarios in which dogs encounter dog-like delivery robots and birds are deterred by bird-like drones. The literature also raises design considerations, calling for animal-aware robots capable of interpreting species-specific behaviours and responding appropriately~\cite{schneiders2024designing}.

Multiple studies highlight that drones often cause fear, stress, or aggression, particularly due to their unpredictable movement patterns, auditory output, and visual appearance. For example, dogs often respond with distress when encountering drones at close range, reacting with avoidance, barking, or aggression~\cite{zamansky2016dog, zamansky2018effects, ahmed2022user, foster2019preliminary, kresnye2021movement}. Wildlife, too, is impacted by drone operations. Behavioural studies show that species such as deer, elephants, birds, and ungulates exhibit heightened sensitivity to drone presence, with responses varying based on altitude, flight path, sound frequency, and species-specific traits~\cite{dorrenbacher2024navigating, chang2017spiders, hoople2020drone, abioye2024mapping, kresnye2021payload, vanvuuren2023ungulate, webber2021digital}. For example, elephants showed increased discomfort in response to drone noise~\cite{abioye2024mapping}, while ungulates reacted more strongly to low-altitude flights and repeated passes~\cite{vanvuuren2023ungulate}. In particular, \citet{rebolo2019drones} reports that more than a quarter of disturbed species belong to IUCN threat categories, strengthening concerns about drone-induced stress in vulnerable populations. The notion of habituation also appears in the literature as a double-edged concept: while familiarity may reduce stress, repeated exposure may still alter animal behaviour in unintended ways~\cite{hirskyj2021reflecting}.

\onlinebadge{Ethnography} \textit{Defensive or uncertain reactions to drones and robots; anecdotal reports raise ethical and legal concerns.}

Aerial systems frequently triggered defensive or aggressive behaviours. Species including alligators, birds of prey, deer, kangaroos, cats, and large mammals such as tigers and bears were seen swatting, chasing, or attacking drones. Some animals, such as lions and moose, showed more cautious or strategic responses. These interactions suggest that drones are often perceived as territorial threats or unfamiliar intruders.

Delivery robots and robotic dogs often caused fear or uncertainty among domestic dogs. In most videos, dogs were seen barking, retreating, or freezing. Robot dogs in particular, especially those that moved suddenly, elicited confusion and distress. A few videos, however, showed neutral or indifferent behaviour, such as dogs calmly observing delivery robots or briefly engaging before losing interest. These cases highlight the variability in animal responses, influenced by prior experience, robot form, and movement patterns.

Reddit posts also highlighted behavioural impacts of autonomous systems on animals, particularly through personal anecdotes and legal interpretations. In the context of aerial systems, two posts sparked a discussion about whether drone-filming of wildlife constitutes harassment. One user, while filming deer and coyotes, questioned whether approaching animals with drones caused meaningful distress. In contrast, another post by a hunting attorney explicitly warned that drone use could be interpreted as unlawful wildlife harassment under state statutes, equating the act with other forms of disturbance such as chasing animals. 

\subsubsection{Ethics and Regulations}\hfill\\
\indent \litbadge{Literature} \textit{Human life prioritised in AV ethics; regulatory focus on physical harm; algorithmic bias concerns.}

Several contributions reveal that public and regulatory opinions often accept the sacrifice of animals in unavoidable collisions, prioritising human life~\cite{awad2020crowdsourcing, steen2024problem, tolmeijer2021machine}. The German Ethics Code~\cite{luetge2017german} similarly formalises this view, stating that systems must be programmed to accept damage to animals or property if doing so prevents injury to humans. The code also acknowledges the constitutional protection of ‘higher animals’ in Germany. For mobile robots, safety is addressed in EN ISO 13482:2014~\cite{ISO13482}, the sole reference standard for personal care robots. While the standard outlines requirements for safe operation near humans, it does not address the safety of bystanders such as pedestrians and animals~\cite{salvini2021iso}. Both are treated as generic \textit{`safety-related objects,'} rendering animals functionally equivalent to static obstacles. More positively, the European Union’s Green Smart Directive calls for smart technologies to adhere to general principles of ecological preservation. This has influenced the programming of garden robots, including robotic lawnmowers, to take greater care in avoiding harm to animals and insects~\cite{hassenzahl2022european}.

Other studies raise broader concerns about how AI systems embed and reinforce value hierarchies. \citet{mancini2023responsible} cautions that AI technologies may unintentionally harm animals not only through system malfunction or limitations, but also through algorithmic bias that reinforces anthropocentric prejudice.

\onlinebadge{Ethnography} \textit{Concerns about exposing pets to robots or leaving them unsupervised in AVs; drone-wildlife regulation.}

In one widely shared video, a dog was seen riding alone in a Tesla, prompting the person filming to ask, \textit{`Is this legal?'} and \textit{`Where is your owner?'}. While the dog appeared calm, the footage triggered questions about accountability, safety, and the appropriateness of leaving animals in AVs unsupervised. 
In one Reddit discussion, a hunting attorney warned against using drones to film or follow wildlife, referencing U.S. state laws that prohibit wildlife harassment. Although the drones discussed were not autonomous, the concerns raised, about disturbance, intent, and legal ambiguity, are relevant to the future use of autonomous aerial systems.

\subsubsection{Accessibility Concerns} \hfill\\
\indent \litbadge{Literature} \textit{Challenges for service animals interacting with AVs and delivery robots; restrictive pet policies.}

Several studies document how delivery robots and AVs confuse or obstruct guide dogs, undermining their effectiveness. In \cite{shin2024delivering}, participants described how unpredictable movements of delivery robots and unfamiliar auditory signals caused disorientation, hesitation, or misdirection in guide dogs. Similar concerns were echoed in interviews with visually impaired users, who reported `stalemate' scenarios in which neither the guide dog nor the robot would yield, leading to pavement blockages and uncertainty~\cite{bhat2022confused}. The problem also extends to vehicle acoustics. Guide dog handlers raised concerns about the near-silent operation of electric and hybrid vehicles, noting that dogs struggle to judge the speed and direction of these vehicles~\cite{hwang2024towards}. Institutional and commercial systems may also fall short. Ride-share services have drawn criticism for not consistently accommodating service animals, in contrast to regulatory requirements applied to traditional taxi services~\cite{bennett2021accessibility}.

\onlinebadge{Ethnography} \textit{Service animal users' concerns about pet policies, public stigma, and animal discomfort with AVs.}

Although no videos depicted accessibility scenarios, Reddit posts offered insights into the experiences of service animal users. One post pointed out Waymo’s strict pet policy, which only allows trained service animals and excludes emotional support animals, highlighting the regulatory and operational distinctions that may affect disabled users. Another post described an individual’s frustration at being criticised online after sharing their first ride with a guide dog, reflecting the ongoing social stigma and misunderstanding faced by service animal users even when policies allow them. A third post offered a more positive experience, praising Waymo's inclusive policy but noting that their service dog appeared uneasy in the absence of a visible driver, initially showing signs of discomfort.

\subsubsection{Urban Disturbance} \hfill\\
\indent \litbadge{Literature} \textit{Robot-triggered animal behaviours causing minor disruptions in shared spaces.}

A final theme identified in the literature concerns secondary disturbances in urban environments. These disturbances, while often minor, reveal how animal responses to robotic presence can ripple out to affect the broader public. Multiple studies report that delivery and service robots can provoke sudden or disruptive behaviours in animals, particularly dogs, leading to unintentional obstructions, slowing pedestrian flow, or creating momentary bottlenecks in shared spaces~\cite{han2024codesign, weinberg2023sharing}. Similarly, \cite{yu2024out} described incidents in which proximity to a robot triggered nervous reactions in dogs, disrupting the quiet of a residential morning and prompting reflections on animals’ right to coexist in public spaces.

\subsection{Expert Interviews}
\label{sec:experts}

The expert interviews extended and contextualised the five thematic categories identified in the scoping review and online ethnography by contributing domain-specific knowledge, while also highlighting cross-cutting design considerations for multispecies coexistence.

\begin{figure*} [ht]
    \centering
\includegraphics[width=0.95\linewidth]{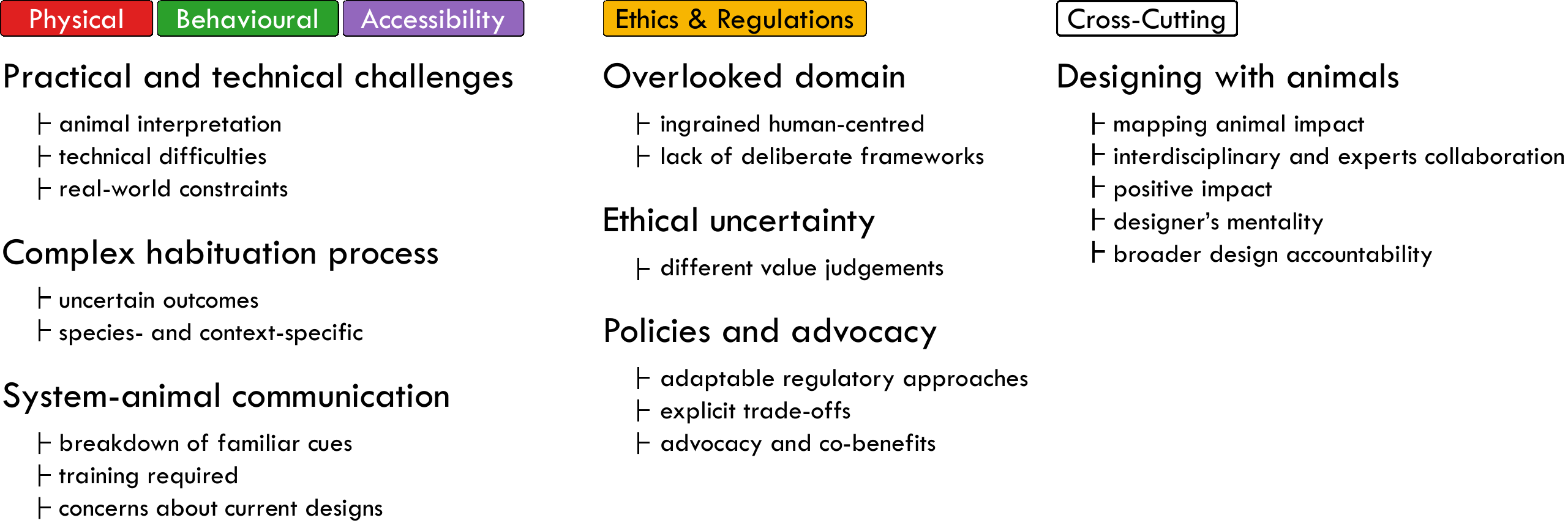}
    \caption{Expert insights are mapped into related thematic categories, with new cross-cutting design directions highlighted for multispecies coexistence. Note: Some insights span multiple categories; none relate directly to Urban Disturbance category.}
    \Description{This figure organises expert insights into three groups. (1) Physical, Behavioural, and Accessibility covers challenges in animal interpretation, habituation, and communication. (2) Ethics & Regulations addresses overlooked domains, ethical uncertainty, and policy needs. (3) Cross-Cutting includes design principles such as mapping animal impact, interdisciplinary collaboration, and broader accountability. No insights map directly to Urban Disturbance.}
    \label{fig:expert-mapping}
\end{figure*}

\paragraph{Practical and technical challenges.}
Experts described how \textbf{animal interpretation} presents unique challenges due to sensory and cognitive differences. One noted, \textit{`We’re biologically different. We try our best, but we must always bear in mind that we’ll never completely know'} (E2). Unlike human participants, animals cannot be interviewed or surveyed, and many methods do not translate (E2, E7). E4 pointed to \textbf{technical difficulties} in predicting animal movement, which is often less predictable than human behaviour. Behavioural differences across species present unique challenges, requiring tailored approaches; for example, kangaroos move unpredictably in bursts, while cassowaries tend to follow set paths. Experts also noted intra-species variation (e.g., a leashed/unleashed dog, a trained service dog) (E6), adding further complexity to detection and decision-making. Animal interactions with autonomous mobility introduces \textbf{real-world constraints} not present in controlled environments (E1). Cost and efficiency also pose challenges: \textit{`You spare the lives of animals, but you lose money because your deliveries will be slower'} (E3).

\paragraph{Complex habituation process.}
Experts agreed that it remains unclear how animals will respond to autonomous systems over time, highlighting \textbf{uncertain outcomes in the habituation process} (E2, E5, E7). Animals may struggle to form accurate expectations of AV behaviour. As one expert noted, \textit{`They’re all cars, but human-driven cars and autonomous vehicles behave very differently'} (E4). There are also examples of habituation; dogs, for example, may become accustomed to delivery robots as the novelty wears off (E1, E2). However, the potential for habituation is \textbf{species- and context-specific} (E2, E5): \textit{`There isn’t an exact line between habituation and not'} (E2).

\paragraph{System-animal communication.}
Experts raised concerns about the \textbf{breakdown of familiar cues} in interactions with AVs. Unlike human drivers, AVs lack behavioural cues such as eye contact or gestures. Experts noted that \textit{`a big part of that mutual communication will just be missing'} (E7), recalling, for example, a guide dog barking at a human driver who stopped short at a crossing. This led to discussions about the \textbf{training required} for service animals (E1, E5, E7, E8); for example, ensuring that a guide dog can recognise an AV as a vehicle they are meant to board (E1). Participants also expressed \textbf{concerns about current designs}, particularly regarding external interface signals that may be ineffective or uncomfortable for animals, such as high-frequency sounds or colours they cannot perceive (E6, E7).

\paragraph{Overlooked domain.}
Experts acknowledged the importance of researching animal interactions, noting \textit{`we knew they are part of the picture'} and that the topic is often overlooked (E6, E8). E3 attributed this oversight in part to \textbf{ingrained human-centred} and anthropocentric perspectives. They also pointed to a \textbf{lack of deliberative frameworks} for representing non-human interests, describing conservation science as good at describing the world, but slow to address the ethical and political implications of acting on behalf of other species.

\paragraph{Ethical uncertainty}
Experts reflected on the \textbf{different value judgements} behind prioritising human lives over animal lives, and among animal species, highlighting that these can be emotional, cultural, or philosophical. Emotionally, one expert admitted, \textit{`I would value the life of a dog more than that of a spider. One is cute and cuddly, and the other is, well, creepy'} (E7). Culturally, practices like culling wild horses to protect native frogs in Australia reflect region-specific value systems (E3, E5). Philosophically, some questioned whether rarity should influence moral weight: \textit{`We have nine billion humans, but there are species of animals with only a few hundred left. So whose life is more important?'} (E2). As one expert cautioned, \textit{`If we are going to employ certain kinds of values or categorisations, then we need to be clear that that’s not science'} (E3), while another added, \textit{`ethical considerations must be informed by data'} (E5).

\paragraph{Policies and advocacy.} 
Strict animal ethics protocols exist in research settings, as highlighted by E1’s reference to a study that reflects on complex ethical review processes in multispecies technology research~\cite{benford2024charting}, and by E5’s mention of institutional SOPs requiring researchers to maintain distance from wildlife. However, experts identified a regulatory gap in autonomous mobility and called for \textbf{adaptable regulatory approaches} (E6, E8). For example, agile standards can evolve faster than formal laws (E6). E8 echoed this view, emphasising that regulation spans laws, standards, and non-binding guidelines, and must keep pace with rapid AI development. 

E3 discussed the potential for \textbf{explicit trade-offs}, arguing that autonomous systems can make decision-making more transparent, since parameters such as animal collision avoidance can be explicitly set in software. They even proposed a hypothetical scenario where \textit{`you could see a button, turn it on or off'} (E3), potentially allowing mobility efficiency at the cost of greater harm to animals. This possibility underscores the need for regulatory oversight (E3, E8). Finally, participants described the \textbf{advocacy and co-benefits} of designing for animal-inclusive futures. Initiatives framed as mutually beneficial for both humans and animals were seen as more persuasive and easier to promote (E1). Several highlighted collaborations between engineering and conservation, where technical and ecological goals align (E4), and pointed to individuals already interacting closely with animals, such as pet owners and veterinarians, as potential advocates (E7).

\paragraph{Designing with animals.}
The design process, experts noted, should begin with \textbf{mapping animal impact}. This requires a deep understanding of animal biology, behaviour, and sensory perception, informed by literature, field observations, and tools such as accelerometers or vocalisation analysis (E1). Experts also highlighted the need for \textbf{interdisciplinary and expert collaboration} (E1–4). Involving animal welfare specialists, veterinarians, and behavioural researchers from the outset improves both animal wellbeing and technical outcomes. One expert reflected, \textit{`We used RGB and thermal cameras, but later found cassowaries aren’t active at night'} (E4). Beyond harm prevention, some experts encouraged designers to strive for \textbf{positive impact}: \textit{`Not every collision has to be bad, if you can design them properly, it could become something positive'} (E1). This might mean reimagining interaction, \textit{`What if an animal could hop on and be carried somewhere, like using it as a little bus or ride?'} (E2). 

More importantly, experts emphasised the importance of the \textbf{designer’s mentality} in addressing animal inclusion (E2, E8). As E8 noted, \textit{`Before we even talk about the design or the technical capabilities robots should have, we need to talk about intent [...] it’s whether we actually want to solve these problems'}. This perspective encourages designers to consider animals as individuals with distinct needs and experiences. As one expert put it, \textit{`A species is not sentient in itself, only the individuals are'} (E3). Finally, some experts also called for \textbf{broader design accountability}, highlighting the long-term ecological implications of autonomous systems. E8 raised sustainability concerns around extractive materials, waste and end-of-life disposal, while E3 highlighted that these systems inevitably participate in ecosystems—carrying not just parcels, but also insects, bacteria, and other beings, and shaping how life responds to their presence.







\section{Discussions}

\subsection{Impacts on Animals Across Species and Contexts (RQ1)}

In response to RQ1, our findings in Section \ref{sec:lit_ethno} show that animals are affected by autonomous systems in different ways. Experts also emphasised the importance of considering the full spectrum of interaction types: \textit{`direct or indirect, cognitive or physical, dyadic or distributed, synchronous or asynchronous.'} (E3). While most current mitigation efforts focus on physical safety (e.g., collision avoidance), behavioural impacts require a more nuanced and animal welfare–oriented lens. In some contexts, animals are already integral to discussions of accessibility, particularly in the case of assistance animals. However, inclusion in autonomous mobility~\cite{locken2021wecare, colley2020towards} requires more than accessible design for humans; it calls for \textit{relational inclusivity} that accounts for human–animal partnerships and the animal’s ability to engage with autonomy on its own terms. Urban disturbances from robot–animal interactions may compound existing accessibility challenges in crowded public spaces. Mobile robots already narrow pedestrian paths and pose barriers for people with mobility impairments~\cite{han2024codesign, bennett2021accessibility}. When animals react unpredictably, these interactions can create additional disturbances~\cite{weinberg2023sharing}. Species-, context-, and system-specific variability may complicate this further. 

The impacts stem not only from the autonomous systems themselves but also from the infrastructure that supports them, such as roadside units or charging stations, which can disrupt animal habitats or routines (as argued by E4). Given the rapid scale of deployment, for example, Waymo reported more than 50,000 paid rides per week~\cite{autoura_apollo}, the frequency and scope of animal encounters with these systems will only increase.

\subsection{Design and Policy Opportunities for Multispecies Coexistence (RQ2)}

In addressing RQ2, our findings highlight the limitations of current design paradigms and regulatory frameworks. As highlighted in the \textit{Ethics \& Regulations} theme (Sections~\ref{sec:lit_ethno} and~\ref{sec:experts}), autonomous systems are largely developed within a human-centred design paradigm, which positions humans as the primary users, beneficiaries, and reference points (e.g.,~\cite{luetge2017german, ISO13482}). However, there is precedent for broader ethical consideration. Wildlife protection laws for drones and more systemic efforts like the European Union’s Green Smart Directive~\cite{hassenzahl2022european}, demonstrate that animal welfare and ecological preservation can be meaningfully incorporated into policy and design guidelines across system types.

From a design perspective, the challenge lies not in the lack of possible approaches, but in the absence of a fundamental mindset shift. While this paper does not argue for a shift to animal-centred design, but rather, the need to recognise animals as relational actors and co-inhabitants of shared environments. This means acknowledging that autonomous systems operate within broader ecological systems, of which humans are only one part. As ~\citet{rosen2025introducing} note, \textit{`a more-than-human turn in design is responding to the role of design in our ecological crises'}, reflecting a growing effort in HCI to expand its scope toward more inclusive, ecologically attuned forms of design. 

\section{Conclusion and Limitations}

The study has several limitations. The scoping review and online ethnography are constrained by our search strategy, which affects the visibility and inclusion of relevant materials, potentially overlooking encounters in under-reported regions or involving less visible species. For example, the animals listed in \autoref{tab:animal_categories} are largely Northern Hemisphere–centric. In contrast, road casualties in Australia often involve wombats, wallabies, and possums~\cite{englefield2020australian}. This underrepresentation may reflect the limited public deployment of AVs and mobile robots in Australian contexts to date. We also sought to include deployments in Asia (e.g., Baidu Apollo), but found no relevant results, likely because large-scale and fully driverless operations only scaled up in 2024~\cite{autoura_apollo}, and information may exist primarily in Chinese and cannot be captured by searches in English. YouTube and Reddit content, which is user-curated, may also skew toward unusual or noteworthy events. Although expert interviews provide valuable domain insights, the sample remains small and selective. 

Despite these limitations, this study identified both direct and indirect ways in which autonomous mobility systems affect animals. It calls for a shift in mindset, one that recognises animals as relational actors and considers their experiences. This work gives visibility to a domain that has been largely overlooked and offers a foundation for future research. Our mapping of expert insights, layered onto findings from literature and real-world encounters (\autoref{fig:expert-mapping}), provides an initial structure for this emerging area and helps surface cross-cutting design directions for multispecies coexistence.

\begin{acks}
This research was supported by the Australian Research Council (ARC) Discovery Project DP220102019, Shared-space interactions between people and autonomous vehicles. The authors thank all experts who generously shared their insights during the study. With permission, we acknowledge the following individuals by name: Anne Quain, Clara Mancini, Daniel Ramp, Debargha Dey, and Mao Shan. Paul Schmidt initially contributed as an expert interviewee and later joined the authorship team to help strengthen the industry perspective of the paper. We also extend our sincere gratitude to the anonymous reviewers for their valuable comments and suggestions.
\end{acks}


\bibliographystyle{ACM-Reference-Format}
\bibliography{references}

\clearpage
\appendix

\section{Scoping Review}
\label{appendix:scoping}

\subsection{ACM Search Log}

\begin{itemize}
  \item \textbf{Search Date:} 19–20 February 2025
  \item \textbf{Publication Years:} 2014–2025
  \item \textbf{Search Within:} Full-text search. Initial title and abstract searches returned few or no relevant results.
  \item \textbf{Search Queries:} See Table~\ref{tab:acm-search-results} for full list of keyword combinations and result counts.
  \item \textbf{Content Type:} \\
  \textit{Included:} Research Articles, Surveys, Short Papers, Columns, Extended Abstracts, Posters, Work in Progress, Opinions (to ensure broad coverage). \\
  \textit{Excluded:} Proceedings Collections, Demonstrations, Courses, Tutorials, Editorials (formats with limited or peripheral research content).
  \item \textbf{Tools}: We used the ACM Digital Library’s advanced search with filters for year and content type. Each query result was exported as an Endnote file and converted to CSV using Publish or Perish\footnote{\url{https://harzing.com/resources/publish-or-perish}}.
  \item \textbf{Note:} We initially attempted to use combined queries with multiple OR clauses (e.g., \texttt{("animal" OR "wildlife") AND ("autonomous vehicle" OR "self-driving car")}), but the ACM advanced search, for some reasons, did not reliably process grouped logical operators at that time. We therefore adopted a simpler strategy using keyword pairs.
\end{itemize}

\begin{table}[ht]
\centering
\footnotesize
\caption{ACM Digital Library keyword search results (2014–2025).}
\begin{tabular}{@{}ll@{}}
\toprule
\textbf{Keywords} & \textbf{Results} \\
\midrule
"animal interaction" AND "delivery robots" & 1 \\
"animals" AND "delivery robots" & 23 \\
"wildlife" AND "delivery robots" & 2 \\
"dogs" AND "delivery robots" & 12 \\
"cats" AND "delivery robots" & 3 \\
"animal interaction" AND "self-driving cars" & 0 \\
"animals" AND "self-driving cars" & 93 \\
"wildlife" AND "self-driving cars" & 13 \\
"animal interaction" AND "autonomous vehicles" & 3 \\
"animals" AND "autonomous vehicles" & 168 \\
"wildlife" AND "autonomous vehicles" & 36 \\
"animal interaction" AND "drones" & 7 \\
"animals" AND "drones" & 318 \\
"wildlife" AND "drones" & 126 \\
\midrule
\textbf{Total} & \textbf{805} \\
\bottomrule
\end{tabular}
\label{tab:acm-search-results}
\end{table}

\subsection{Google Scholar Search Log}
\begin{itemize}
  \item \textbf{Search Date:} 19 February 2025
  \item \textbf{Publication Years:} 2014–2025
  \item \textbf{Search Within:} Anywhere in the paper.
  \item \textbf{Search Queries:} See Table~\ref{tab:gs-search-results} for full list of keyword combinations and result counts.
  \item \textbf{Tools}: The search was conducted using the Publish or Perish interface, and results were saved as a CSV file.
  \item \textbf{Note:} Initial queries using broad terms such as "animals" or "wildlife" returned an unmanageably large number of publications, many of which were unrelated to interaction or mobility contexts. To better capture relevant studies, we narrowed the queries to include phrases such as "animal interaction" and "wildlife interaction".
\end{itemize}

\begin{table}[ht]
\centering
\footnotesize
\caption{Google Scholar keyword search eesults (2014–2025). }
\begin{tabular}{@{}ll@{}}
\toprule
\textbf{Keywords} & \textbf{Results} \\
\midrule
"animal interaction" AND "self driving cars" & 24 \\
"animal interaction" AND "autonomous vehicles" & 53 \\
"animal interaction" AND "delivery robots" & 10 \\
"animal interaction" AND "delivery drones" & 3 \\
"animal interaction" AND "drones" & 223 \\
"wildlife interaction" AND "self driving cars" & 3 \\
"wildlife interaction" AND "autonomous vehicles" & 5 \\
"wildlife interaction" AND "delivery robots" & 0 \\
"wildlife interaction" AND "delivery drones" & 0 \\
"wildlife interaction" AND "drones" & 105 \\
\midrule
\textbf{Total} & \textbf{426} \\ 
\bottomrule
\end{tabular}
\label{tab:gs-search-results}
\end{table}


\begin{table*}[htbp]
\centering
\footnotesize
\caption{Overview of papers where animals featured in broader debates on ethics, policy, or directions for ACI (n=36).}
\renewcommand{\arraystretch}{1.4} 
\begin{tabular}{p{3.5cm} p{0.5cm} p{10cm}}
\toprule
\textbf{Authors} & \textbf{Year} & \textbf{Paper Title} \\
\midrule
\citet{hermann2024ai} & 2024 & AI-Powered Animal Recognition and Tracking via Drone-Based Thermal Imaging: A User-Centric Mobile Application \\
\citet{steen2024problem}& 2024 & The Problem with the Trolley Problem and the Need for Systems Thinking \\
\citet{dorrenbacher2024navigating}& 2024 & Navigating the Paradox: Challenges of Designing Technology for Nonhumans \\
\citet{akridge2024bus}& 2024 & “The Bus is Nothing Without Us”: Making Visible the Labor of Bus Operators amid the Ongoing Push Towards Transit Automation \\
\citet{han2024codesign}& 2024 & Co-design Accessible Public Robots: Insights from People with Mobility Disability, Robotic Practitioners and Their Collaborations \\
\citet{schneiders2024designing}& 2024 & Designing Multispecies Worlds for Robots, Cats, and Humans \\
\citet{yu2024out}& 2024 & Out of Place Robot in the Wild: Envisioning Urban Robot Contextual Adaptability Challenges Through a Design Probe \\
\citet{shin2024delivering}& 2024 & Delivering the Future: Understanding User Perceptions of Delivery Robots \\
\citet{abioye2024mapping}& 2024 & Mapping Safe Zones for Co-located Human-UAV Interaction \\
\citet{li2024creating}& 2024 & Creating with More-Than-Humans \\
\citet{mancini2023responsible}& 2023 & Responsible ACI: Expanding the Influence of Animal-Computer Interaction \\
\citet{joshi2023autonomous}& 2023 & Autonomous Vehicle and External Road User Interfaces: Mapping of Standards Gaps and Opportunities \\
\citet{yaghoubisharif2022mobile}& 2022 & HeadsUp: Mobile Collision Warnings through Ultrasound Doppler Sensing \\
\citet{ahmed2022user}&2022 & User-Centred Design Methods in Animal-Centred Design: A Systematic Review \\
\citet{hassenzahl2022european}&2022 & European Union’s Green Smart Directive or How Resource-Conscious Smart Systems Saved the World \\
\citet{hermann2022police} & 2022 & User-Defined Gesture and Voice Control in Human-Drone Interaction for Police Operations \\
\citet{hirskyj2021reflecting} & 2021 & Reflecting on Methods in Animal Computer Interaction: Novelty Effect and Habituation \\
\citet{thrift2021killer}& 2021 & Killer Cities \\
\citet{tolmeijer2021machine}& 2021 & Implementations in Machine Ethics: A Survey \\
\citet{salvini2021iso}& 2021 & On the Safety of Mobile Robots Serving in Public Spaces: Identifying Gaps in EN ISO 13482:2014 and Calling for a New Standard \\
\citet{bennett2021accessibility}& 2021 & Accessibility and The Crowded Sidewalk: Micromobility’s Impact on Public Space \\
\citet{webber2021digital}& 2021 & Digital Technologies in Nature \\
\citet{kresnye2021payload}& 2021 & Payload Drones and ACI: Drone Navigation System Prototype \\
\citet{awad2020crowdsourcing} & 2020 & Crowdsourcing Moral Machines \\
\citet{nijholt2020virtual}& 2020 & Virtual and Augmented Reality Animals in Smart and Playful Cities \\
\citet{jin2020acoussist}& 2020 & Acoussist: An Acoustic Assisting Tool for People with Visual Impairments to Cross Uncontrolled Streets \\
\citet{hoople2020drone}& 2020 & Drone Use Case Studies \\
\citet{parks2019mediating}& 2019 & Mediating Animal Infrastructure Relations \\
\citet{owens2019rewilding}& 2019 & Rewilding Cities \\
\citet{foster2019preliminary}& 2019 & Preliminary Evaluation of Dog-Drone Technological Interfaces: Challenges and Opportunities \\
\citet{bendel2018towards}& 2018 & Towards animal-friendly machines \\
\citet{luetge2017german}& 2017 & The German ethics code for automated and connected driving \\
\citet{chang2017spiders}& 2017 & “Spiders in the Sky”: User Perceptions of Drones, Privacy, and Security \\
\citet{vaataja2014animal}& 2014 & Animal Welfare as a Design Goal in Technology Mediated Human-Animal Interaction \\
\citet{zamansky2016dog}& 2016 & Dog-Drone Interactions: Towards an ACI Perspective \\
\citet{kalik2007automotive} & 2007 & Automotive Turing Test \\
\bottomrule
\end{tabular}
\end{table*}

\begin{table*}[ht]
\centering
\footnotesize
\caption{Summary of empirical studies on animal interaction with autonomous mobility systems (n=9).}
\renewcommand{\arraystretch}{1.4} 
\begin{tabular}{p{2.5cm} p{0.6cm} p{4.8cm} p{5.5cm}}
\toprule
\textbf{Author(s)} & \textbf{Year} & \textbf{Paper Title} & \textbf{Study Type} \\
\midrule
\citet{hwang2024towards} & 2024 & Towards Robotic Companions: Understanding Handler-Guide Dog Interactions for Informed Guide Dog Robot Design & Semi-structured interviews and observation sessions with 23 dog guide handlers and 5 trainers. \\
\citet{vanvuuren2023ungulate} & 2023 & Ungulate Responses and Habituation to Unmanned Aerial Vehicles in Africa's Savanna & Examined behavioral responses of free-ranging ungulate species in Namibia to different in-flight UAV models. \\
\citet{vaataja2023exploring} & 2023 & Exploring Dogs’ Reactions when Encountering Delivery Robots in Urban Environment & Online survey of 212 dog owners reporting behavior of dogs encountering grocery delivery robots. \\
\citet{weinberg2023sharing} & 2023 & Sharing the Sidewalk: Observing Delivery Robot Interactions with Pedestrians during a Pilot in Pittsburgh, PA & Ethnographic observations and intercept interviews with residents during a public sidewalk robot pilot. \\
\citet{bhat2022confused} & 2022 & `I was Confused by It; It was Confused by Me' – Exploring the Experiences of People with Visual Impairments around Mobile Service Robots & Interviews with 17 people with visual impairments about experiences with vacuum, delivery, and drone robots \\
\citet{kresnye2021movement} & 2021 & Movement Patterns as Enrichment: Exploratory Canine-Drone Interaction Pilot Study & Species-centric evaluation of drone flying patterns and distances for enrichment purposes. \\
\citet{rebolo2019drones} & 2019 & Drones as a Threat to Wildlife: YouTube Complements Science in Providing Evidence about Their Effect & Combined scientific literature with Internet sources (e.g., YouTube) to examine drone effects on wildlife. \\
\citet{zamansky2018effects} & 2018 & Effects of Anxiety on Canine Movement in Dog-Robot Interactions & Exploratory study of 20 dogs, comparing those with deprivation syndrome to healthy controls in dog-robot interaction scenarios. \\
\citet{rasmussen2021wildlife} & 2011 & Wildlife Conservation at a Garden Level: The Effect of Robotic Lawn Mowers on European Hedgehogs & Described and quantified effects of robotic lawn mowers; tested 18 mowers in collisions with dead hedgehogs. \\
\bottomrule
\end{tabular}
\label{tab:animal_robot_studies}
\end{table*}

\section{Online Ethnography}
\label{appendix:ethnography}


\begin{itemize}
  \item \textbf{Search Date:} 12 March 2025
  \item \textbf{Platforms:} YouTube and Reddit (accessed without signing in)
  \item \textbf{System-related keywords:} self driving cars, autonomous vehicles, Tesla, Waymo, delivery robots, drones
  \item \textbf{Animal-related keywords:} wildlife, animals, animal interaction, pets, cats, dogs, deers
  \item \textbf{Note:} Searches were conducted using combinations of system-related and animal-related keywords.
    \item \textbf{Total:} YouTube (N=60), Reddit (N=29)
\end{itemize}


\begin{table*}[ht]
\centering
\footnotesize
\caption{Reddit threads analysed in online ethnography.}
\begin{tabular}{@{}p{7.5cm}p{6.5cm}@{}}
\toprule
\textbf{Title} & \textbf{Reddit Link} \\
\midrule
Please Don't Vandalize or Bring Dogs into Waymos & \url{https://www.reddit.com/r/SelfDrivingCars/comments/1fsli1f/please_dont_vandalize_or_bring_dogs_into_waymos/} \\
Guide dog rider discriminated against & \url{https://www.reddit.com/r/service_dogs/comments/1gf7cu3/i_posted_a_video_of_me_and_my_guide_dog_riding/} \\
Waymo scares my dog & \url{https://www.reddit.com/r/service_dogs/comments/1c99bne/waymo_scares_my_dog/} \\
Drone-filming wildlife = harassment? & \url{https://www.reddit.com/r/drones/comments/1i73j4f/at_what_point_is_dronefilming_wildlife_considered/} \\
PSA: Stop Using Drones to Harass Wildlife & \url{https://www.reddit.com/r/Hunting/comments/4xe6j1/psa_from_hunting_attorney_stop_using_drones_to/} \\
Tesla's Full Self-Driving v13 stops for Cat Crossing Road & \url{https://www.reddit.com/r/SelfDrivingCars/comments/1hkhib2/teslas_full_selfdriving_v13_stops_for_cat/} \\
What's going on with the 'Kangaroo Problem' & \url{https://www.reddit.com/r/SelfDrivingCars/comments/uioske/whats_going_on_with_the_kangaroo_problem/} \\
Tesla driver hits a dog, claims they were on Full Self-Driving Beta & \url{https://www.reddit.com/r/electricvehicles/comments/145hbyr/tesla_driver_hits_a_dog_claims_they_were_on_full/} \\
Update from Waymo spokesperson on the dog killed & \url{https://www.reddit.com/r/SelfDrivingCars/comments/142xe34/update_from_waymo_spokesperson_on_the_dog_that/} \\
Deer: A self-driving car's worst nightmare & \url{https://www.reddit.com/r/SelfDrivingCars/comments/37kbt4/deer_a_self_driving_cars_worst_nightmare/} \\
Waymo stops for small critters & \url{https://www.reddit.com/r/waymo/comments/1hiesyi/waymo_stops_for_small_critters/} \\
\bottomrule
\end{tabular}
\label{tab:reddit-discussions}
\end{table*}

\begin{table*}[ht]
\centering
\footnotesize
\caption{YouTube videos analysed in online ethnography. Note: Video titles have been standardised for consistency. Emojis and hashtags were removed from original titles to ensure compatibility with LaTeX formatting and improve readability.}
\begin{tabular}{@{}p{6.5cm}p{7cm}@{}}
\toprule
\textbf{Video Title} & \textbf{YouTube Link} \\
\midrule
Huge alligator tries to catch a drone in the Amazon river! & \url{https://www.youtube.com/shorts/PnnpVsOMBQ4} \\
Alligator vs Drone & \url{https://www.youtube.com/shorts/mHCznxR4geg} \\
Animals Destroying Drones Caught On Camera & \url{https://www.youtube.com/watch?v=zgFzxdAkHUQ} \\
Animals vs. Drones & \url{https://www.youtube.com/watch?v=OgmVQLz3-Ms} \\
Drone vs Cat & \url{https://www.youtube.com/shorts/CIi1BTDcd_Q} \\
Harley meets a delivery robot & \url{https://www.youtube.com/shorts/wAZeIY2Y_S8} \\
Bella spooked by Starships delivery robot & \url{https://www.youtube.com/shorts/JbH9iSsEvjk} \\
Confronting a robot & \url{https://www.youtube.com/shorts/rJkUWoHsnps} \\
Chick-Fil-A AI delivery robot & \url{https://www.youtube.com/shorts/51dSetmjF2M} \\
Delivery Robot scares dog & \url{https://www.youtube.com/watch?v=JNpYh_yRPQ0} \\
Moi greets Starship delivery robot & \url{https://www.youtube.com/shorts/Fka2TyOOh2Y} \\
Encounter with Levi the delivery robot & \url{https://www.youtube.com/shorts/UVeEi-5bKsM} \\
Who is more scared? Dog or robot? & \url{https://www.youtube.com/shorts/rDwvqdnN7dE} \\
Food robot in Finland & \url{https://www.youtube.com/shorts/UeDcLOKP4wg} \\
Dog scared of drone & \url{https://www.youtube.com/shorts/CO1ORzjj2P8} \\
Drone and dog funny reels & \url{https://www.youtube.com/shorts/F1Zdp4sbZSU} \\
Drone and eagle interaction & \url{https://www.youtube.com/shorts/9ZPof5XCUlw} \\
Retired police horse with Waymo & \url{https://www.youtube.com/watch?v=4eKr02qrvNQ} \\
Dogs meet a robot dog & \url{https://www.youtube.com/shorts/Si4GlwC5sIQ} \\
What is this dog? Robot dog reaction & \url{https://www.youtube.com/shorts/Geux8Alp53I} \\
Robot Dog greets dogs & \url{https://www.youtube.com/shorts/LT454AThOq4} \\
Golden retriever meets robot dog & \url{https://www.youtube.com/shorts/LpSqBwQ4Qi8} \\
Dog doesn't like robot dog & \url{https://www.youtube.com/shorts/4yRjJYBXnlQ} \\
Tesla's Autopilot Hits Animals (2024) & \url{https://www.youtube.com/watch?v=ngFDWJrbe6Y} \\
Tesla Model 3 Hits Deer & \url{https://www.youtube.com/watch?v=FeQPaWFyiPE} \\
Tesla senses bears on highway & \url{https://www.youtube.com/shorts/9l2sf4v-13E} \\
Tesla avoids wild animal & \url{https://www.youtube.com/shorts/T5GP2HNGUgE} \\
Tesla avoids deer at night & \url{https://www.youtube.com/watch?v=pWJoe8hwu_I} \\
Tesla detects deer & \url{https://www.youtube.com/shorts/VVwUBuLLJpI} \\
Tesla avoids 2 deer – close call & \url{https://www.youtube.com/watch?v=BvObcjBJHTI} \\
Tesla FSD V12 Deer Test & \url{https://www.youtube.com/watch?v=1TUUtfV841s} \\
Autopilot and Animals – Close Encounter & \url{https://www.youtube.com/watch?v=R1fQcRfyVFU} \\
Tesla sees deer, reacts quickly & \url{https://www.youtube.com/shorts/3942SERH8JM} \\
Tesla brakes for deer, VW hits & \url{https://www.youtube.com/watch?v=TNrxv6Gh8-s} \\
Waymo detects dog in SF & \url{https://www.youtube.com/watch?v=AQAxJZPWn0s} \\
Waymo detects running dog & \url{https://www.youtube.com/watch?v=7mJJ9WM1Rf8} \\
Tesla FSD v12.5.4 – Misses Animals? & \url{https://www.youtube.com/watch?v=ncPNKl66P00} \\
Tesla FSD Beta vs Animals & \url{https://www.youtube.com/watch?v=-y9zgAd63JA} \\
Tesla Self Driving Dog & \url{https://www.youtube.com/shorts/B1nfB9mQk2I} \\
\bottomrule
\end{tabular}
\label{tab:youtube-videos}
\end{table*}


\begin{table*}[ht]
\centering
\footnotesize
\caption{Expert profiles: Experts E1–E5 have direct experience with animals; E6–E8 offer adjacent perspectives from autonomous systems and sociotechnical fields.}
\begin{tabular}{p{0.8cm}p{3cm}p{9cm}}
\toprule
\textbf{ID} & \textbf{Expertise Area(s)} & \textbf{Profile Summary} \\
\midrule
E1 & HRI, animal welfare & \textbf{HRI/HCI Researcher.} Investigates how autonomous systems, including robots, impact human and non-human relationships. \\
E2 & ACI & \textbf{ACI Professor.} Leads an ACI lab and authored the ACI manifesto to establish animal-centred computing as a discipline. \\
E3 & wildlife conservation,\newline road ecology & \textbf{Conservation Scientist.} Focuses on animal behavioural responses to vehicles, sensory modalities, and mitigating anthropogenic impacts. \\
E4 & connected and automated\newline vehicles (CAV) & \textbf{CAV Researcher.} Works on V2X communication and CAV safety. Recent projects include cassowary detection and road safety. \\
E5 & animal welfare,\newline veterinary ethics & \textbf{Veterinarian and Specialist in Animal Welfare Science, Ethics and Law.} Research focuses on animal welfare science, human–animal interaction, and veterinary ethics. \\
\addlinespace
\hline
\addlinespace
E6 & autonomous vehicles, \newline systems engineering & \textbf{Industry Expert.} Experience in consumer robotics, robotaxis, and automotive safety. Focuses on systems integration, software, and the societal role of AVs. \\
E7 & human factors,\newline smart mobility & \textbf{Mobility Researcher.} Studies human practice in AV systems, last-mile mobility, urban tech, and smart infrastructure. \\
E8 & robotics, semiotics & \textbf{HRI Researcher.} Explores robotic technologies in everyday spaces and how meaning emerges through design, culture, and socio-material practices. \\
\bottomrule
\end{tabular}
\label{tab:expert_profiles}
\end{table*}

\end{document}